\documentclass[9pt,technote]{IEEEtran}%%%% As of March 2017, [siggraph] is no longer used. Please use sigconf (above) for SIGGRAPH conferences.

\usepackage{graphicx}
\usepackage{microtype}
\usepackage{hyperref}
\usepackage{amssymb}
\usepackage[table,xcdraw]{xcolor}
\usepackage{natbib}
\makeatletter
\let\old@ps@headings\ps@headings
\let\old@ps@IEEEtitlepagestyle\ps@IEEEtitlepagestyle
\def\confheader#1{%
  \def\ps@headings{%
    \old@ps@headings%
    \def\@oddhead{\strut\hfill#1\hfill\strut}%
    \def\@evenhead{\strut\hfill#1\hfill\strut}%
  }%
  \def\ps@IEEEtitlepagestyle{%
    \old@ps@IEEEtitlepagestyle%
    \def\@oddhead{\strut\hfill#1\hfill\strut}%
    \def\@evenhead{\strut\hfill#1\hfill\strut}%
  }%
  \ps@headings%
}
\makeatother

\confheader{%
  2018 (preprint)
}

\begin{document}

\title{Improving Trace Link Recommendation by Using Non-Isotropic Distances and Combinations}


%\author{\IEEEauthorblockN{
%Christof Tinnes\IEEEauthorrefmark{2},
%Andreas Biesdorf\IEEEauthorrefmark{2},
%Uwe Hohenstein\IEEEauthorrefmark{2}}
%\IEEEauthorblockA{\IEEEauthorrefmark{2}Siemens AG - Technology, Otto-Hahn-Ring 6, M{\"u}nchen 81739, Germany \\
%\{christof.tinnes\}@siemens.com}}

\author{
\IEEEauthorblockN{Christof Tinnes\IEEEauthorrefmark{1}}
%\vspace{0.05in}
\IEEEauthorblockA{\IEEEauthorrefmark{1}Siemens AG. \emph{christof.tinnes@siemens.com}}
%\vspace{0.05in}
}

%\author{Christof Tinnes}
%\email{christof.tinnes@siemens.com}
%\affiliation{%
%  \institution{Siemens AG}
%  \streetaddress{Otto-Hahn-Ring 6}
%  \city{München}
% \state{Germany}
%  \postcode{81739}
%}
%
\maketitle              % typeset the header of the contribution

\begin{abstract}
The existence of trace links between artifacts of the software development life cycle can improve the efficiency of many activities during software development, maintenance and operations. Unfortunately, the creation and maintenance of trace links is time-consuming and error-prone. Research efforts have been spent to automatically compute trace links and lately gained momentum, e.g., due to the availability of powerful tools in the area of natural language processing. In this paper, we report on some observations that we made during studying non-linear similarity measures for computing trace links. We argue, that taking a geometric viewpoint on semantic similarity can be helpful for future traceability research. We evaluated our observations on a dataset of four open source projects and two industrial projects. We furthermore point out that our findings are more general and can build the basis for other information retrieval problems as well.

% \keywords{First keyword  \and Second keyword \and Another keyword.}
\end{abstract}
\section{Introduction}
Creating and maintaining links between related software development artifacts supports many tasks in software engineering. For example, links between source code and design decisions could improve source code understanding and inhibit design erosion. Links between requirements and source code can be used for bug localization, since end user problem descriptions are usually formulated in the language of the requirements and therefore bug reports and corresponding requirements should have a high similarity. Links between requirements and source code can further be used for software reuse \cite{Bouillon2013}. Having links between tickets in an issue management system and duplicate tickets can avoid unnecessary efforts for bug fixing. There are certainly many more applications of links between artifacts in the software development lifecycle. \par
Unfortunately, the creation and maintenance of these links is time consuming. We think the development of tools to support software practitioners in creating and maintaining trace links is worthwhile \cite{Tinnes2019}. 

Motivated by our industrial experience, we considered three concrete real world challenges described by the following three tasks: \par
\begin{bfseries}Task 1:\end{bfseries} For a given source code commit, we want to recommend a link to a corresponding ticket or issue in an issue management system. Some workflows already support the linking of source code commits to corresponding issues in an issue management system, e.g., by using pull requests but many workflows still lack this linking. In these cases, we believe that recommendation of trace links can improve the traceability in the project. \par
\begin{bfseries}Task 2:\end{bfseries} A typical problem in industrial projects is the detection of duplicate tickets. Often, software defects are discovered independently, for example, at different testing levels. Detecting duplicate candidates at ticket creation or during bug scrub meetings helps consolidating similar tickets and reduces bug fixing efforts.  \par 
\begin{bfseries}Task 3:\end{bfseries} We trace the summary of a ticket to the description of the same ticket. Of course, these links are given in practise but this task resembles the retrieval of tickets based on input queries. We consider the summary of the ticket as the search query and the description as the content of the ticket. We mainly used this task as a test task for our pilot study to pick approaches that work well for trace link creation.

As has been shown by Antoniol et al. \cite{Antoniol2002}, information retrieval techniques can be employed to create trace links.
For a given artifact, the most similar artifacts in a given set are retrieved via information retrieval techniques and can be recommended as candidate trace links. Hence, trace link recommendation is reduced to the problem of computing artifact similarities. The same applies to detecting duplicate tickets. The semantic similarity of two duplicates should be high compared to other tickets. So basically all three tasks can be considered as information retrieval problems and similarity measures can be used to retrieve candidates.
\par
There are several possibilities to compute document similarities. Basically there are set-theoretic models (e.g., boolean retrieval), probabilistic models (e.g., Latent Dirichlet Allocation (LDA) \cite{Blei2003}), and vector space models (e.g., Latent Semantic Indexing (LSI) \cite{Deerwester90indexingby}) for information retrieval.
\par
Based on the vectors space model (VSM) approach, we want to motivate the development of geometric models for information retrieval. In the VSM approach to information retrieval, the query (i.e., the information need) and the objects to be retrieved are mapped into a common vector space. A similarity measure is then used to retrieve the most similar objects for the input query. More formally, the idea of the VSM is to represent natural language artifacts in a (semi-) metric space $(X, d)$. The distance $d(x,y)$ should be small for documents $x, y$ (or words) that are semantically similar and large for documents that are not similar. Since Mikolov's seminal paper in 2013 \cite{Mikolov2013}, Word2Vec and also other Word Embedding techniques (e.g., FastText \cite{bojanowski2016enriching} or Glove \cite{pennington2014glove}) have been applied to many information retrieval tasks and distance computation \cite{Goth2016}. Word embeddings map words into a real-valued vector space. They bridge the gap between Harris' distributional hypothesis \cite{harris54} and the research on distributed representations \cite{Hinton1986}. Word2Vec has been related to the mutual information measure \cite{Levy2014} that has been known for a long time in information retrieval.\par

In this work, we want to take a geometric viewpoint on distributed representations. It is not the purpose of this study, to provide algorithms that perform better than current state-of-the algorithms for creating trace-links. Instead, we want to show, that the geometry of the space of document representations is relevant for computing the similarity of two artifacts. This can build the basis for development of better trace link recommendation techniques in the future.

\section{Approach and Research Questions}
For our study of information retrieval similarity measures for trace link recommendation, duplicate detection and ticket retrieval, we considered more concretely the following seven datasets: \\
Traceability (TR) datasets, i.e., source code commits to ticket links:
\begin{enumerate}
    \item TR-HADOOP - Apache Hadoop project
    \item TR-MYFACES - Apache MyFaces project
    \item TR-FELIX - Apache Felix project
    \item TR-AMELIE - Industrial project within Siemens called AMELIE
\end{enumerate}
Duplicate (DP) datasets, i.e., duplicate links between two tickets/issues in an Issue Management System:
\begin{enumerate}
  \setcounter{enumi}{4}
    \item DP-JENKINS - The Jenkins Project
    \item DP-SKY - An industrial project at Sky
\end{enumerate}
Our information retrieval (IR) ``test problem'' dataset, i.e., tracing between summary and description of a ticket/issue.
\begin{enumerate}
  \setcounter{enumi}{6}
    \item IR-FELIX - Apache felix project
\end{enumerate}

For the commit-to-ticket retrieval tasks, we used the ticket id that has been included in the corresponding source code commit messages for these projects. We cleaned the datasets by removing all source code commits without Jira ticket id or tickets without description. For the duplicate tasks, we selected only the tickets that have a Jira "duplicates" link.

There are several approaches and models available to compute document similarities. We focus on creating document vectors out of word vectors.
These document vectors are efficient to compute and the same model can be used for two different artifacts types (in the case of trace links). We initially compared Word Mover's Distance (WMD) \cite{Kusner2015}, Doc2Vec \cite{Le2014} and the cosine distance of the averaged Word2Vec word vectors. WMD and Doc2Vec did not perform well for our task and furthermore, for our evaluations, we need to compute approximately $10^8$ similarities and WMD took too much computation time. We also trained a Siamese LSTM network, similar to the approach used for sentence similarity in \cite{Mueller2016}, to compute document similarities. On our datasets we also obtained poor results. \par
For this study, we focused on document embeddings that are averaged word embeddings. It is fast, easy to implement, and we initially obtained good results, without any sophisticated adaption.\par
\begin{figure}[ht]
			\centering
			\includegraphics[width=\columnwidth]{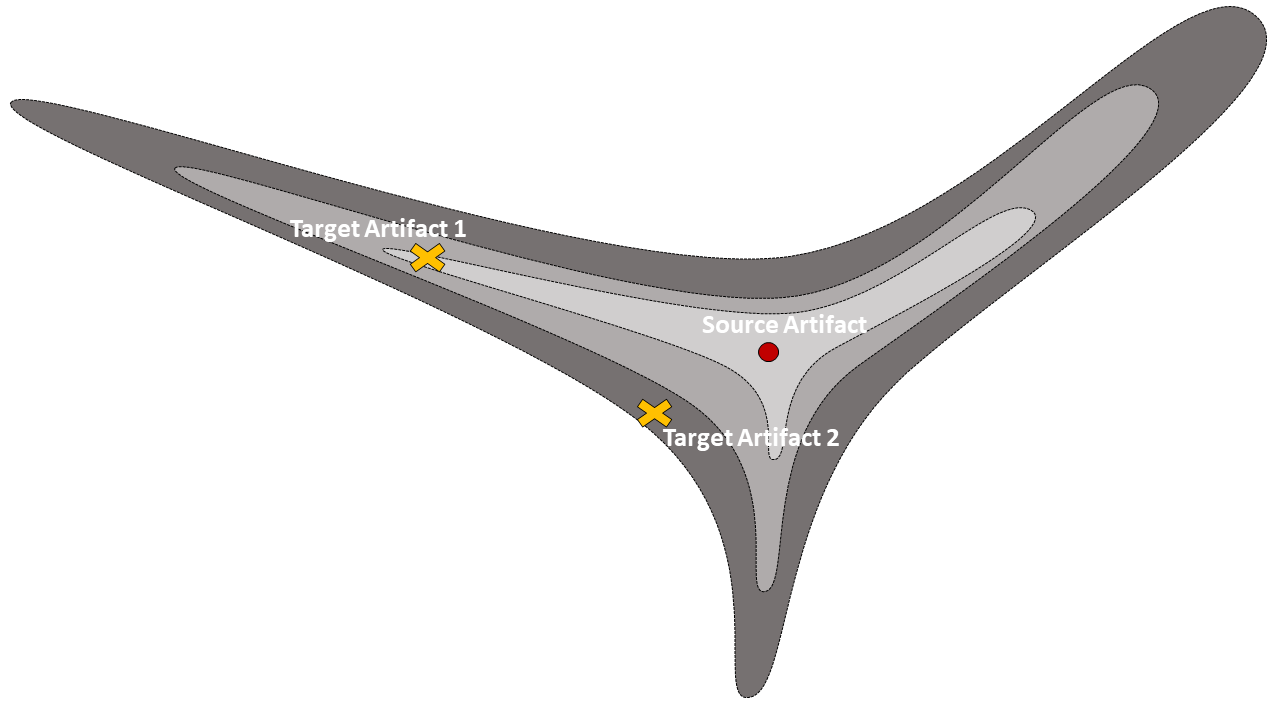}
			\caption{Even though target artifact 2 is closer to the source artifact using euclidean distance, target artifact 1 is closer to the source artifact in the distance indicated by the level set.}
			\label{fig:non-linear-distances}
\end{figure}

%Of course, there are many more approaches to create document representations out of word representations and to compute document similarities and we leave a broader comparison for future research.\par
Cosine similarity assumes that the document embeddings live in a linear space and that the distance is isotropic, i.e., independent of the direction. To the best of our knowledge, it has not been shown that this assumption of isotropy is true (for word embeddings). 
It is thus very likely that the ``semantic'' structures and the metrics for semantic similarity are rather non-linear and not all dimensions of the feature space (i.e., the document embeddings) contribute equally to the semantic distance. We argue that the semantics of the documents is tightly coupled to geometric properties of some (unknown) geometric structure (e.g., Riemannian manifolds). Our documents vectors are embeddings from this geometric structure to $\mathbb{R}^n$, where $n$ is the dimension of the document embeddings. To compute some distance (or similarity), it is therefore not correct to use the euclidean or cosine distance. Instead, one has to also embed the distance measure from geometric structure to $\mathbb{R}^n$. As indicated in Figure \ref{fig:non-linear-distances} even though one target artifact might be closer to a source artifact using the euclidean distance, this might not be true for this embedded distance measure. \par
Unfortunately, we do not know the geometric structure. We therefore need to find (or approximate) the distance measure based on our existing data. \par
Taken the intrinsic manifold structures of the search spaces into account has already been studied more generally for information retrieval problems \cite{Zhou2004}. 
Unlike the manifold ranking approach in \cite{Zhou2004}, we propose to learn a non-linear semi-distance based on labeled data. Heuristically motivated by Arnold's representation theorem \cite{Kolmogorov1956}, we propose a two layer neural network architecture. For the first layer, we use ReLu activation functions, for the second layer Tanh activation functions. For the output neuron we use the Sigmoid activation function, since we want to interpret the result as the probability that there is a link between the two artifacts. We furthermore use Dropout layers. We experimented also with other two layer architectures but this configuration performed well throughout all of our datasets. %Of course, more sophisticated network architectures might perform better but we want to keep things simple to qualitatively compare the cosine distance with the learned non-linear semi-distance. 
For two artifacts, we compute the difference of their document embeddings and feed them into the network.
We train the models on our datasets for the three tasks mentioned earlier. For example, for Jira tickets that are not duplicates, we expect a similarity of $0$ and for duplicate Jira tickets, the similarity is expected to be $1$. The loss function is then set to be the binary cross-entropy. We optimize the network using Adam optimizer. Hyperparameteres are fixed for all datasets. They were chosen to perform well on our test problem (IR-FELIX) and also show sound accuracy through all tasks and datasets.

We then want to answer the following questions:

\smallskip
\begin{itemize}
    \item \begin{bfseries} Research Question 1 - \end{bfseries}  What is the influence of the  word embeddings models on our tasks? \par
    \begin{itemize}
        \item \begin{bfseries} Research Question 1.1 - \end{bfseries} Can general purpose models be used and how do they compare to models specifically trained on the given dataset?\par
        \item \begin{bfseries} Research Question 1.2 - \end{bfseries} Can multiple models for the word embeddings be combined to increase the performance?\par
    \end{itemize}
    \item \begin{bfseries} Research Question 2 - \end{bfseries}  How does the learned distance measure compare to the cosine approach? \par
    \begin{itemize}
        \item \begin{bfseries} Research Question 2.1 - \end{bfseries} How do both approaches compare qualitatively?
        \item \begin{bfseries} Research Question 2.2 - \end{bfseries} How does a combination of cosine and neural network distances perform on our tasks?
        \item \begin{bfseries} Research Question 2.3 - \end{bfseries} Can neural networks trained on one dataset be reused for other datasets?
    \end{itemize}
\end{itemize}

Our main goal is the comparison of the learned non-linear distance and the cosine distance. In order to have results that are independent on the training of the Word2Vec model itself, we first aimed at understanding the influence of the Word2Vec models on our tasks (Research Question 1). Especially we would like to use the same ``general purpose'' models throughout all tasks and all datasets. To further improve the performance of the ``general purpose'', we experimented with multiple model combinations.
Research Question 2.1. is the central question of this paper. Since we observed (see below) a ``crossing'' of the Acc@k curves for the non-linear distance approach and the cosine approach, we wanted to see if the combination of cosine and non-linear distance approach even performs better than any single approach (Research Question 2.2.). Research Question 2.3. is a kind of cross check for our hypothesis. Since we claim that the neural network learns a distance metric specific to the language used in the project, we would assume that the trained network does not perform well on other datasets from other projects.\par
\smallskip

To evaluate our approach, we compute a matrix of the distances between all source- and target artifacts. The distance is either the cosine distance defined by
\begin{equation}
    d(x,y) = 1 - \frac{<x,y>}{|x||y|},
\end{equation} or the result of applying the trained neural network to the difference vector of source- and target artifact, respectively. 
We then compute the percentage of correctly retrieved target artifacts (implemented tickets, duplicate ticket, ticket description) for the given source artifacts (source code commits, tickets with duplicates, ticket summary). We vary the number of retrieved documents, in which our target artifact should be contained. We refer to this measures in the following by Accuracy@k, though in the language of information retrieval measures it would rather be the mean recall@k.
By ``combination'' in the research questions, we refer to computing a distance matrix by averaging the distance matrices to be combined.

\section{Results}
We look at the following measures to answer our research questions:
\begin{enumerate}
\item Accuracy@k (acc@k): The relative amount of correct target artifacts when k artifacts are retrieved.
\item AUC: The area under the Accuracy@k curve.
\end{enumerate}
For our tasks, the Accuracy@k measure tells us how good the algorithm performs, when $k$ recommendations are presented to a user. 
Higher values of the AUC are a good indicator but not a guarantee for a better performance of the approach.
We are mostly interested in low values of $k$, since we only want to make a few recommendations. An approach with a low AUC might still perform well for
low values of $k$, since the acc@k curve could stagnate for higher values of $k$.

%The computation of the distance matrices for the open source projects took 14h on Google Colaboratory with Python 3 Google Compute Engine Backend(TPU) and 25GB RAM. The distance matrices for the industrial projects were computed locally and took less then 10 minutes on a Intel Core i5, 8GB RAM machine (they were smaller in size compared to the OSS projects).\par
We will list here only our main findings. For details, we suggest to take a look at our code\footnote{https://github.com/chtinnes/nonlinear-distance-for-traceabiliy} and the datasets from the open source projects.

To answer \begin{bfseries} research question 1.1\end{bfseries}, we trained a Word2Vec model using the Python gensim library. As training corpus, we used the source- and target artifacts. 

\begin{table*}[t]
\caption{AUC measure for all datasets and the different models used.}
\label{tbl:AUC-models}
\centering
\begin{tabular}{|l|l|l|l|l|l|l|l|l|}
\hline
\rowcolor[HTML]{9B9B9B} 
{\color[HTML]{FFFFFF} Dataset}                & {\color[HTML]{FFFFFF} Cos-Self} & {\color[HTML]{FFFFFF} Cos-SO}         & {\color[HTML]{FFFFFF} Cos-GN} & {\color[HTML]{FFFFFF} Cos-2M} & {\color[HTML]{FFFFFF} NL-Self} & {\color[HTML]{FFFFFF} NL-SO} & {\color[HTML]{FFFFFF} NL-GN} & {\color[HTML]{FFFFFF} NL-2M}          \\ \hline
TR-HADOOP                            & 0.676                           & \textbf{0.787}                        & 0.676                         & 0.762                         & 0.820                          & 0.927                        & {\color[HTML]{000000} 0.924} & {\color[HTML]{000000} \textbf{0.951}} \\ \hline
\rowcolor[HTML]{C0C0C0} 
{\color[HTML]{000000} IR-FELIX} & {\color[HTML]{000000} 0.792}    & {\color[HTML]{000000} \textbf{0.886}} & {\color[HTML]{000000} 0.818}  & {\color[HTML]{000000} 0.878}  & {\color[HTML]{000000} 0.904}   & {\color[HTML]{000000} 0.973} & {\color[HTML]{000000} 0.964} & {\color[HTML]{000000} \textbf{0.985}} \\ \hline
TR-MYFACES                           & 0.641                           & \textbf{0.774}                        & 0.671                         & 0.745                         & 0.768                          & 0.918                        & {\color[HTML]{000000} 0.909} & {\color[HTML]{000000} \textbf{0.943}} \\ \hline
\rowcolor[HTML]{C0C0C0} 
TR-FELIX                            & 0.699                           & \textbf{0.868}                        & 0.763                         & 0.852                         & 0.778                          & 0.948                        & {\color[HTML]{000000} 0.933} & {\color[HTML]{000000} \textbf{0.964}} \\ \hline
TR-AMELIE                            & 0.681                           & 0.634                                 & \textbf{0.699}                & 0.672                         & 0.628                          & 0.893                        & {\color[HTML]{000000} 0.883} & {\color[HTML]{000000} \textbf{0.957}} \\ \hline
\rowcolor[HTML]{C0C0C0} 
DP-JENKINS                            & 0.800                           & \textbf{0.877}                        & 0.825                         & 0.869                         & 0.849                          & 0.959                        & {\color[HTML]{000000} 0.949} & {\color[HTML]{000000} \textbf{0.970}} \\ \hline
DP-SKY                           & 0.788                           & \textbf{0.818}                                 & 0.800                & 0.817                         & 0.865                          & 0.921                        & {\color[HTML]{000000} 0.930} & {\color[HTML]{000000} \textbf{0.945}} \\ \hline
\end{tabular}
\end{table*}

\begin{figure}[ht]
			\centering
			\includegraphics[width=\columnwidth]{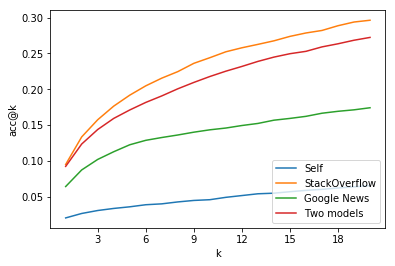}
			\caption{Project=Apache Hadoop, distance=cosine; The Accuracy@k curves or the Apache Hadoop dataset and the Word2Vec model trained on the dataset itself, the Google News Word2Vec model and the StackOverflow Word2Vec model. The combination of two models is also shown. Similarity has been computed using cosine distance.}
			\label{fig:rq1-cosine}
\end{figure}

\begin{figure}[ht]
			\centering
			\includegraphics[width=\columnwidth]{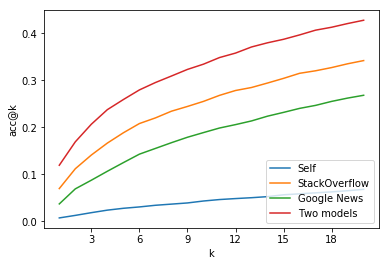}
			\caption{Project=Apache Hadoop, distance=non-linear; The Accuracy@k curves or the Apache Hadoop dataset and the Word2Vec model trained on the dataset itself, the Google News Word2Vec model and the StackOverflow Word2Vec model. The combination of two models is also shown. Similarity has been computed using the non-linear distance.}
			\label{fig:rq1-nn}
\end{figure}

We have set the hyperparameters fixed so that they perform reasonable for all of our datasets. We compared these trained Word2Vec models to the StackOverflow (SO) model from \cite{Efstathiou2018}, and the original 
Word2Vec model from \cite{Mikolov2013} trained on the Google News corpus. These two models are publicly available and can be used with the gensim library. We did this comparison for the cosine distances as well as for our computed non-linear distances.
When looking at the plots (see Figure \ref{fig:rq1-cosine} and Figure \ref{fig:rq1-nn}) and also at the AUC measure (Table \ref{tbl:AUC-models}), we can clearly see that the general purpose models perform better.
The Accuracy@k curves for the other datasets look qualitatively similar. We also optimized the Word2Vec parameters for a specific dataset which gave us results comparable to the StackOverflow model.\par
Regarding \begin{bfseries}research question 1.2\end{bfseries}, for the non-linear distance measure, we can see that combining the SO Word2Vec model and the Google News Word2Vec model give us better results than for any single model.
For the cosine distance, this is not true. The Accuracy@k curves for the StackOverflow model mostly dominate the combined model curve.

\begin{figure}[ht]
			\centering
			\includegraphics[width=\columnwidth]{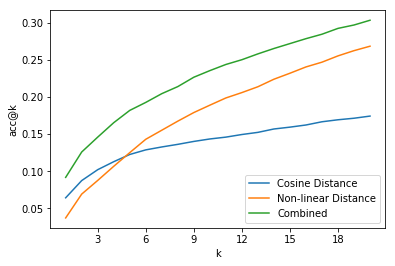}
			\caption{Project=Apache Hadoop, distance=cosine vs. non-linear;The Accuracy@k curve for the non-linear distance approach dominates the acc@k curve for the cosine distance approach. Here shown for the Apache Hadoop Traceability dataset and the StackOverflow Word2Vec model.}
			\label{fig:crossing}
\end{figure}

To answer \begin{bfseries} research question 2.1\end{bfseries}, we compared the cosine distance and the non-linear distance approach qualitatively. When looking at the Accuracy@k curves, we can see from Table \ref{tbl:crossing} that there is always (for all of our datasets and for the SO, Google News Word2Vec input vector models) a $K_{dom} \geq 0$ for which the the trained non-linear distance approach outperforms the cosine distance approach (see Figure \ref{fig:crossing} for an example). More interestingly, as can be seen from Table \ref{tbl:crossing}, this $k$ is usually small and in a range which is interesting for recommendations, on average $k = 6.38$. We can also see from Table \ref{tbl:crossing} that, except for the duplicate tasks, the acc@k curves for the  non-linear distance measure completely dominate the cosine distance approach in the case we combined the two Word2Vec models. The AUC measures in Table \ref{tbl:auc-rq2} support these observations.

\begin{table*}[t]
\caption{AUC measures for the Cosine, Non-linear(NL), and the combined distances for the StackOverflow(SO), Google News(GN) and the combined Word2Vec models(2M).}
\begin{tabular}{|l|l|l|l|l|l|l|l|l|l|}
\hline
\rowcolor[HTML]{9B9B9B} 
{\color[HTML]{FFFFFF} \textbf{Dataset}}       & {\color[HTML]{FFFFFF} \textbf{Cosine-SO}} & {\color[HTML]{FFFFFF} \textbf{NL-SO}} & {\color[HTML]{FFFFFF} \textbf{Combined-SO}} & {\color[HTML]{FFFFFF} \textbf{Cosine-GN}} & {\color[HTML]{FFFFFF} \textbf{NL-GN}} & {\color[HTML]{FFFFFF} \textbf{Combined-GN}} & {\color[HTML]{FFFFFF} \textbf{Cosine-2M}} & {\color[HTML]{FFFFFF} \textbf{NL-2M}} & {\color[HTML]{FFFFFF} \textbf{Combined-2M}} \\ \hline
TR-HADOOP                            & 0.787                                     & 0.927                                 & 0.917                                       & 0.676                                        & 0.924                                    & 0.927                                          & 0.762                                     & \textbf{0.951}                        & 0.939                                       \\ \hline
\rowcolor[HTML]{C0C0C0} 
{\color[HTML]{000000} IR-FELIX} & {\color[HTML]{000000} 0.886}              & {\color[HTML]{000000} 0.973}          & {\color[HTML]{000000} 0.966}                & {\color[HTML]{000000} 0.818}                 & {\color[HTML]{000000} 0.973}             & {\color[HTML]{000000} 0.954}                   & {\color[HTML]{000000} 0.878}              & {\color[HTML]{000000} \textbf{0.985}} & 0.978                                       \\ \hline
TR-MYFACES                           & 0.774                                     & 0.918                                 & 0.904                                       & 0.671                                        & 0.909                                    & 0.918                                          & 0.745                                     & \textbf{0.943}                        & 0.927                                       \\ \hline
\rowcolor[HTML]{C0C0C0} 
TR-FELIX                            & 0.868                                     & 0.948                                 & 0.943                                       & 0.763                                        & 0.933                                    & 0.948                                          & 0.852                                     & \textbf{0.964}                        & 0.958                                       \\ \hline
TR-AMELIE                            & 0.634                                     & 0.893                                 & 0.700                                       & 0.699                                        & 0.883                                    & 0.893                                          & 0.672                                     & \textbf{0.957}                        & 0.758                                       \\ \hline
\rowcolor[HTML]{C0C0C0} 
DP-JENKINS                            & 0.877                                     & 0.959                                 & 0.954                                       & 0.825                                        & 0.949                                    & 0.942                                          & 0.869                                     & \textbf{0.970}                        & 0.963                                       \\ \hline
DP-SKY                            & 0.818                                     & 0.921                                 & 0.918                                       & 0.800                                        & 0.930                                    & 0.922                                          & 0.817                                     & \textbf{0.945}                        & 0.936                                       \\ \hline
\end{tabular}
\label{tbl:auc-rq2}
\end{table*}

\begin{table}[]
\caption{Minimal k ($K_{dom}$), for which the non-linear distance dominates the cosine distance.}
\resizebox{\columnwidth}{!}{%
\begin{tabular}{|l|l|l|}
\hline
\rowcolor[HTML]{9B9B9B} 
{\color[HTML]{FFFFFF} \textbf{Dataset}}  & {\color[HTML]{FFFFFF} \textbf{Word2Vec model}} & {\color[HTML]{FFFFFF} \textbf{$K_{dom}$}} \\ \hline
TR-HADOOP                       & StackOverflow                                  & 12                                     \\ \hline
\rowcolor[HTML]{C0C0C0} 
{\color[HTML]{000000} TR-HADOOP} & {\color[HTML]{000000} Google News}             & {\color[HTML]{000000} 1}               \\ \hline
TR-HADOOP                       & StackOverflow + Google News                    & 0                                      \\ \hline
\rowcolor[HTML]{C0C0C0} 
IR-FELIX                   & StackOverflow                                  & 8                                      \\ \hline
IR-FELIX                   & Google News                                    & 0                                      \\ \hline
\rowcolor[HTML]{C0C0C0} 
IR-FELIX                   & StackOverflow + Google News                    & 0                                      \\ \hline
TR-MYFACES                      & StackOverflow                                  & 2                                      \\ \hline
\rowcolor[HTML]{C0C0C0} 
TR-MYFACES                      & Google News                                    & 0                                      \\ \hline
TR-MYFACES                      & StackOverflow + Google News                    & 0                                      \\ \hline
\rowcolor[HTML]{C0C0C0} 
TR-FELIX                       & StackOverflow                                  & 5                                      \\ \hline
TR-FELIX                       & Google News                                    & 4                                      \\ \hline
\rowcolor[HTML]{C0C0C0} 
TR-FELIX                       & StackOverflow + Google News                    & 0                                      \\ \hline
TR-AMELIE                       & StackOverflow                                  & 0                                      \\ \hline
\rowcolor[HTML]{C0C0C0} 
TR-AMELIE                       & Google News                                    & 0                                      \\ \hline
TR-AMELIE                       & StackOverflow + Google News                    & 0                                      \\ \hline
\rowcolor[HTML]{C0C0C0} 
DP-JENKINS                       & StackOverflow                                  & 17                                      \\ \hline
DP-JENKINS                       & Google News                                    & 8                                      \\ \hline
\rowcolor[HTML]{C0C0C0} 
DP-JENKINS                       & StackOverflow + Google News                    & 3                                      \\ \hline
DP-SKY                       & StackOverflow                                  & 49                                      \\ \hline
\rowcolor[HTML]{C0C0C0} 
DP-SKY                       & Google News                                    & 17                                      \\ \hline
DP-SKY                       & StackOverflow + Google News                    & 8                                      \\ \hline
\end{tabular}
}
\label{tbl:crossing}
\end{table}

To answer \begin{bfseries} research question 2.2\end{bfseries}, we combined the distance matrix for the cosine distance approach and the distance matrix for the non-linear distance approach. Regarding Accuracy@k,
for small values of $k$, the combined approach outperforms the cosine distance as well as the trained non-linear distance approaches. Except for the AMELIE traceability dataset and the Sky dataset with StackOverflow Word2Vec model, the Accuracy@1 (see Table \ref{tbl:rq22-acc1}) is highest for the combined approach. For larger values of $k$ the non-linear distance crosses the combined approach for some $K_{cross} \footnote{The smallest k, for which the non-linear distance crosses the combined distance approach}$. On average this crossing happens for $K_{cross} = 43.14$.
%We can also see from our results that this dominance for $k < K_{cross}$ is negligible in the case of combining the StackOverflow and the Google News model except for the Jira duplicates tasks.

\begin{table*}[t]
\caption{The Accuracy@1 for the datasets and the different approaches.}
\begin{tabular}{|l|l|l|l|l|l|l|l|l|l|}
\hline
\rowcolor[HTML]{9B9B9B} 
{\color[HTML]{FFFFFF} \textbf{Dataset}}       & {\color[HTML]{FFFFFF} \textbf{Cosine-SO}} & {\color[HTML]{FFFFFF} \textbf{NL-SO}} & {\color[HTML]{FFFFFF} \textbf{Combined-SO}} & {\color[HTML]{FFFFFF} \textbf{Cosine-GN}} & {\color[HTML]{FFFFFF} \textbf{NL-GN}} & {\color[HTML]{FFFFFF} \textbf{Combined-GN}} & {\color[HTML]{FFFFFF} \textbf{Cosine-2M}} & {\color[HTML]{FFFFFF} \textbf{NL-2M}} & {\color[HTML]{FFFFFF} \textbf{Combined-2M}} \\ \hline
TR-HADOOP                            & 0.046                                     & 0.020                                 & \textbf{0.056}                              & 0.021                                        & 0.013                                    & \textbf{0.035}                                 & 0.036                                     & 0.049                                 & \textbf{0.056}                              \\ \hline
\rowcolor[HTML]{C0C0C0} 
{\color[HTML]{000000} IR-FELIX} & {\color[HTML]{000000} 0.209}              & {\color[HTML]{000000} 0.168}          & {\color[HTML]{000000} \textbf{0.217}}       & {\color[HTML]{000000} 0.135}                 & {\color[HTML]{000000} 0.138}             & {\color[HTML]{000000} \textbf{0.151}}          & {\color[HTML]{000000} 0.210}              & {\color[HTML]{000000} 0.231}          & \textbf{0.241}                              \\ \hline
TR-MYFACES                           & 0.061                                     & 0.050                                 & \textbf{0.073}                              & 0.034                                        & 0.041                                    & \textbf{0.054}                                 & 0.044                                     & 0.087                                 & \textbf{0.087}                              \\ \hline
\rowcolor[HTML]{C0C0C0} 
TR-FELIX                            & 0.095                                     & 0.070                                 & \textbf{0.109}                              & 0.064                                        & 0.037                                    & \textbf{0.092}                                 & 0.092                                     & 0.119                                 & \textbf{0.124}                              \\ \hline
TR-AMELIE                            & 0.077                                     & \textbf{0.256}                        & 0.103                                       & 0.154                                        & \textbf{0.256}                           & 0.231                                          & 0.128                                     & \textbf{0.385}                        & 0.179                                       \\ \hline
\rowcolor[HTML]{C0C0C0} 
DP-JENKINS                            & 0.189                                     & 0.130                                 & \textbf{0.200}                              & 0.164                                        & 0.131                                    & \textbf{0.178}                                 & 0.189                                     & 0.177                                 & \textbf{0.207}                              \\ \hline
DP-SKY                            & \textbf{0.125}                                     & 0.043                        & 0.121                                       & 0.130                                        & 0.072                           & \textbf{0.134}                                          & 0.136                                     & 0.102                         & \textbf{0.143}                                       \\ \hline
\end{tabular}
\label{tbl:rq22-acc1}
\end{table*}

\begin{table}[]
\label{tbl:rq22-crossing}
\caption{Minimal k ($K_{cross}$), for which the non-linear distance approach dominates the combined distance approach.}
\begin{tabular}{|l|l|l|}
\hline
\rowcolor[HTML]{9B9B9B} 
{\color[HTML]{FFFFFF} \textbf{Dataset}}  & {\color[HTML]{FFFFFF} \textbf{Word2Vec model}} & {\color[HTML]{FFFFFF} \textbf{$K_{cross}$}} \\ \hline
TR-HADOOP                       & StackOverflow                                  & 72                                     \\ \hline
\rowcolor[HTML]{C0C0C0} 
{\color[HTML]{000000} TR-HADOOP} & {\color[HTML]{000000} Google News}             & {\color[HTML]{000000} 35}              \\ \hline
TR-HADOOP                       & StackOverflow + Google News                    & 3                                      \\ \hline
\rowcolor[HTML]{C0C0C0} 
IR-FELIX                   & StackOverflow                                  & 53                                     \\ \hline
IR-FELIX                   & Google News                                    & 3                                      \\ \hline
\rowcolor[HTML]{C0C0C0} 
IR-FELIX                   & StackOverflow + Google News                    & 2                                      \\ \hline
TR-MYFACES                      & StackOverflow                                  & 18                                     \\ \hline
\rowcolor[HTML]{C0C0C0} 
TR-MYFACES                      & Google News                                    & 20                                     \\ \hline
TR-MYFACES                      & StackOverflow + Google News                    & 1                                      \\ \hline
\rowcolor[HTML]{C0C0C0} 
TR-FELIX                       & StackOverflow                                  & 104                                    \\ \hline
TR-FELIX                       & Google News                                    & 76                                     \\ \hline
\rowcolor[HTML]{C0C0C0} 
TR-FELIX                       & StackOverflow + Google News                    & 2                                      \\ \hline
TR-AMELIE                       & StackOverflow                                  & 0                                      \\ \hline
\rowcolor[HTML]{C0C0C0} 
TR-AMELIE                       & Google News                                    & 0                                      \\ \hline
TR-AMELIE                       & StackOverflow + Google News                    & 0                                      \\ \hline
\rowcolor[HTML]{C0C0C0} 
DP-JENKINS                       & StackOverflow                                  & 60                                    \\ \hline
DP-JENKINS                       & Google News                                    & 52                                     \\ \hline
\rowcolor[HTML]{C0C0C0} 
DP-JENKINS                       & StackOverflow + Google News                    & 20                                      \\ \hline
DP-SKY                       & StackOverflow                                  & 223                                      \\ \hline
\rowcolor[HTML]{C0C0C0} 
DP-SKY                       & Google News                                    & 104                                      \\ \hline
DP-SKY                       & StackOverflow + Google News                    & 58                                      \\ \hline
\end{tabular}
\end{table}

\begin{figure}[ht]
			\centering
			\includegraphics[width=\columnwidth]{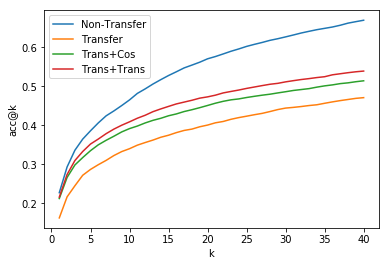}
			\caption{Project=Apache Felix, distance=non-linear; The Accuracy@k curves for the non-linear distance approach for the Apache Felix dataset and classifiers trained on Apache MyFaces and Apache Hadoop datasets. ``Transfer'' means that the classifier has been trained on a different dataset.}
			\label{fig:rq23}
\end{figure}

Regarding \begin{bfseries} research question 2.3\end{bfseries}, we can see from Table \ref{tbl:AUC-transfer} and the Accuracy@k curves (see Figure \ref{fig:rq23} for example) that the models trained on a different dataset perform significantly worse. Combining the distance matrices of classifiers trained on two different datasets or combining the matrix of one classifier with the distance matrix from the cosine distance approach improves the accuracies (except for the Apache Felix Tracability dataset with classifier trained on the Apache Felix Ticket retrieval dataset). But even for the combined approaches, the accuracies are still significantly lower than the ones from the non-transferred approach.

\begin{table*}[t]
\label{tbl:AUC-transfer}
\caption{AUC measures for the transfer approach.}
\begin{tabular}{|l|l|l|l|l|l|l|}
\hline
\rowcolor[HTML]{9B9B9B} 
{\color[HTML]{FFFFFF} \textbf{Dataset}}  & {\color[HTML]{FFFFFF} \textbf{Transfer-DS-1}} & {\color[HTML]{FFFFFF} \textbf{Transfer-DS-2}} & {\color[HTML]{FFFFFF} \textbf{Non-Transfer}} & {\color[HTML]{FFFFFF} \textbf{Transfer}} & {\color[HTML]{FFFFFF} \textbf{Trans+Cos}} & {\color[HTML]{FFFFFF} \textbf{Trans+Trans}} \\ \hline
TR-HADOOP                       & IR-FELIX                        & TR-MYFACES                           & 0.951                                        & 0.826                                    & 0.815                                     & \textbf{0.831}                              \\ \hline
\rowcolor[HTML]{C0C0C0} 
{\color[HTML]{000000} TR-HADOOP} & {\color[HTML]{000000} IR-FELIX} & {\color[HTML]{000000} TR-FELIX}     & {\color[HTML]{000000} 0.951}                 & {\color[HTML]{000000} 0.826}             & {\color[HTML]{000000} 0.815}              & {\color[HTML]{000000} \textbf{0.828}}       \\ \hline
TR-HADOOP                       & IR-FELIX                        & TR-FELIX                            & 0.951                                        & \textbf{0.826}                           & 0.815                                     & 0.811                                       \\ \hline
\rowcolor[HTML]{C0C0C0} 
TR-HADOOP                       & TR-MYFACES                           & TR-FELIX                            & 0.951                                        & 0.789                                    & 0.801                                     & \textbf{0.816}                              \\ \hline
TR-HADOOP                       & TR-MYFACES                           & TR-FELIX                            & 0.951                                        & 0.789                                    & \textbf{0.801}                            & 0.786                                       \\ \hline
\rowcolor[HTML]{C0C0C0} 
TR-HADOOP                       & TR-FELIX                            & TR-FELIX                            & 0.951                                        & 0.800                                    & \textbf{0.803}                            & 0.797                                       \\ \hline
\end{tabular}
\end{table*}

\section{Discussion}
%RQ1.1
Word embeddings map words into a vector space over $\mathbb{R}$. These embeddings carry some semantics such that words which are close to each other tend to be semantically similar. Different embedding models, trained on different corpora lead to different semantics and words can have different semantics, depending on their context. The StackOverflow model performs better in our case, since it has been trained on a software engineering corpus, whereas the Google News models is more ``general purpose''. We would assume therefore that the Word2Vec models trained on the datasets of our study should perform better on our retrieval tasks. This is not the case since the corpora we used to train the models were comparatively small (compared to StackOverflow or GoogleNews). We also made the observation that for small corpora, we need some fine tuning of the hyperparameters for the Word2Vec model training. \par
%RQ 1.2: why does combination of models improves results for NN but not for Cosine? 
%Our NN models are trained to the domain already. There distances are kind of aligned. Averaging the distance from two models is averaging these already good distances.
%For the cosine distance, bad distances might be complete outliers. The will impair the overall distance measure.
%RQ2.1
We have observed that the Accuracy@k curves for the non-linear approach have a higher slope for small values of k but for single Word2Vec models and small k, the cosine approach performs better (``The non-linear acc@k curve crosses the cosine acc@k curve''). 
We cannot completely explain this behaviour yet but it might be due to the fact that for small k, the source and target artifact vectors are very close to each other. For small euclidean distances the ``linear approximation'' might be especially good. The neural network might not be trained to be as good as the cosine approach for these euclidean distances. This would also be supported by the fact that for the approach using two Word2Vec models, the non-linear approach is overall better (except for the duplicate task). Note that we already made the observation that the combination of models works especially well for the non-linear distance approach, and thus it is not very surprising that the dominance of the non-linear approach is stronger for the case of combining the distances for the Google News and StackOverflow Word2Vec models. \par
%RQ2.2
Accordingly, in the case of just using one single Word2Vec model, we could improve the performance for small k by combining the distance matrices of the cosine and non-linear case, i.e., making a compromise between both approaches. It could also be interpreted as a kind of boosting, which is often applied in machine learning. For the combined Word2Vec model case, this dominance of the combined distances for $k < K_{cross}$ might be negligible since our non-linear models already perform well (i.e., better than the cosine approach). For the industrial AMELIE dataset, we cannot make this observations. We consider this as an outlier, since the dataset is very small compared to the other datasets. \par
%RQ2.3
We think, the geometric structures (especially the distance) are related to the semantics. It is therefore not very surprising that the non-linear distance highly depends on the dataset. We can observe again a boosting effect when combining multiple distance matrices but we are not able to get close to the performance of the non-linear distance trained specifically for the dataset. For the Apache Felix traceability dataset with the classifier trained on the Apache ticket retrieval dataset, the transfer learning approach works well. This confirms the hypothesis, since the project and therefore the semantics are the same for both datasets. \par

In general, we can say that it is possible to interpret our results from the geometric viewpoint though some effects have to be investigated further.
\section{Related Work}
 
Since the seminal paper of Antoniol et al. from 2002 \cite{Antoniol2002}, there have been many publications that use information retrieval techniques for establishing trace links (e.g., between source code and documentation or between test cases and requirements).
For example, Asuncion et al. \cite{Asuncion2010} used topic modelling for a semantic categorization of artifacts to support traceability. Shouichi Nagano et al. \cite{Nagano2012} trained a Bayes classifier for trace link recovery.
They parsed the source code and transformed identifier keywords into comment keywords before they fed the data into the classifier.
Yet et al. \cite{Ye2016} apply Learning-To-Rank (LTR) to a bug localization problem and investigate the influence of several features on the ranking task.
These approaches are orthogonal to our results and for future work, our approach could be extended by taking abstract syntax trees and multiple features into account.

Rahman et al. \cite{Rahman2018} conducted a case study on the influence of the similarity measure on a project recommendation task and a bug localization task.
They found that the model choice has a significant impact on the performance and that it also depends on artifact types.

Jin Guo et al. \cite{Guo2017} propose a deep learning approach to trace link generation and also take semantics and domain knowledge into account. As in our work, they used word embeddings as input to their models. Contrary to our work, they used Recurrent Neural Networks to learn artifact semantics, whereas we only consider semantics that is contained in the average of all words in a document.

For the duplicate detection task, there is also a lot of existing work. To just mention a few, Jalbert and Weimer \cite{jalbert08} showed that development cost can be reduced by filtering duplicate bug reports. They trained a classifier based on a combination of categorical and textual features and graph clustering algorithms to identify duplicates. Hindle et al.\cite{Hindle:2016:CAT:2911378.2911428} illustrated that adding contextual information about software-quality attributes allowed the classifiers to decide more efficiently if two bug reports are duplicates or not. They build upon Sun et al. \cite{sun2011} method using an extended version of BM25F for textual similarity.

In contrast to the existing work, we argue that the semantics of the artifacts might be highly related to the geometry of the space which represent our documents. We furthermore provide some concrete results on how a good similarity measure can be constructed out of word embeddings, i.e., learning a non-linear distance measure for each project, and combining the distances for multiple word embedding models.

% statistical analysis of word embedding distributions \cite{Senel2018}

\section{Threads to Validity}

%Conclusion validity
With the exception of the AMELIE dataset, we made all of our observations throughout all tasks and datasets. Since the AMELIE dataset had significantly less tickets, we think that it could be considered to be an outlier here. Furthermore, most of our observations still hold true for the AMELIE dataset.\par

%Internal validity
We fixed hyperparameters for the Word2Vec model training as well as for the neural network training and used the same text preprocessing for all tasks and datasets. We only varied what was necessary to answer the research questions.
The hyperparameters could be better for one dataset than for another, but we have chosen the parameters to lead to stable results.
We applied the trained neural network to compute the distances for all the artifacts in our dataset. Therefore, there is an overlap of the tickets
the network has been trained on and the tickets we are finally using for our evaluation. We did also perform a split of our datasets and obtained qualitatively the
same results though quantitatively we observed differences, which might be caused since we had less data to train the network on.\par

%Construct validity
Of course, there are many more possibilities to compute a non-linear distance measure. Furthermore, our measure does not satisfy the triangle inequality and is therefore only a semi-distance. Our aim was to qualitatively compare non-linear distances to the cosine distance and not to find the ``optimal'' distance for our tasks. This is another reason, why we have chosen a simple two-layer neural network architecture instead of more sophisticated deep learning (e.g., LSTM) approaches. 
Furthermore, other variables (e.g., preprocessing, hyperparameters, word embedding technology, combination of distances, document vectors) could be modified and their effect on the performance of the distance measures be evaluated. \par

% External Validity: 
We made our observations based on three different tasks and seven different datasets. We are confident that our results (i.e., non-linear distance measures with combined multiple general purpose Word2Vec models for the input vectors outperform simple cosine distances) can be generalized to other datasets and could even be generalized to other information retrieval tasks.

\section{Conclusion and Outlook}
In our viewpoint on semantic similarities, artifacts are points on a high-dimensional manifold. Semantic similarity can be measured by a distance measure on this manifold.
Word embeddings learn representations in a vector space and are trained on a specific corpus. When creating document vectors out of word vectors in a different or specific context,
the computed numerical embeddings together with a cosine distance might not fully catch the semantic structures for this context and the given dimension.
We therefore tried to recover the intrinsic semantic structures by learning a non-linear distance measure and therefore ``correct'' the inaccuracies of a context-free approach like just averaging word vectors. Furthermore, different input word embeddings (in our case SO model and GoogleNews model) correspond to different semantic viewpoints on the topic and we belief that a combination of multiple distance measures could improve the accuracy of the approach. 

In this study, we compared the cosine distance approach and a non-linear (semi-)distance approach for document embeddings for three tasks in software engineering.
We also studied the influence of the word embeddings model on the distance measures. Our results show, that non-linear measures are indeed superior to the simple cosine approach, especially when multiple general purpose word embedding distances are combined. This is also in accordance with our geometric viewpoint on semantic similarities. As usual for applications of machine learning, there are many things that could be modified and many possible combinations. Therefore, we left out many questions that could be of interest, for example, 
\begin{itemize}
    \item What is the influence of the preprocessing?
    \item Can we obtain similar results for other word embeddings technologies?
    \item What is the influence of the method used to compute the document embeddings out of word embeddings (e.g., we could also have used Doc2VecC \cite{Chen17aa} or TF-IDF)?
    \item What is the influence of the hyperparameters for the Word2Vec models as well as the trained neural network classifier?
\end{itemize}

For future research, we also want to take a geometric viewpoint on more sophisticated approaches, e.g., using Recurrent Neural Networks as in Guo's work \cite{Guo2017}. 
In the case of deep learning, where for each word the embeddings are directly fed into the network, the document embedding vector space (and it is by far not obvious how to construct this space) is of much higher dimension, which of course might be a better approximation to the intrinsic similarity, but drops the advantages of the dimensional reduction word embeddings give us.

We also left practical consideration out of scope for this study. When looking at our results, we can see that the accuracies are partially low (e.g., Acc@1 $\sim$ 0.15). We compared here only the textual similarities. The recommendations can be significantly improved by taking other features, like quality attribute information (as in \cite{Hindle:2016:CAT:2911378.2911428}), dates, component names, author, parse trees (as in \cite{Nagano2012}) and so on into account and ranking the results using Learning-To-Rank techniques as in \cite{Ye2016}.

To support (or reject) the geometric viewpoint proposed here, other distance measures known from manifold learning (e.g., ISOMAP) should be investigated and the actual geometric properties of the manifolds need to be studied thoroughly.

\def\bibfont{\footnotesize}
\bibliography{geometricdistance}

\end{document}